# The Ehrenfest's Paradox and Radial Electric Field in Quasi-Neutral Tokamak Plasma

**Romannikov A.**

*ITER Agency, Russian Research Center «Kurchatov Institute», pl. Kurchatova I., 1, Moscow 123182, Russia*

*a.romannikov@iterrf.ru*

**Abstract**

A relation between physical consequences of the so-called Ehrenfest's Paradox and the radial electric field $E_r(r)$ in the classical quasi-neutral tokamak plasma is shown. Basic author's approach to the relativistic nature of the tokamak $E_r(r)$ has been described in [1]. The experiment which can resolve the Ehrenfest's Paradox is presented.

**Introduction**

More than 100 years ago, this relativistic paradox was presented in [2] for the first time. Later, it was named the Ehrenfest's Paradox. Detailed historical, physical and geometrical description of the Ehrenfest's Paradox can be found in [3] and references therein. For our experimental purposes, let's present the Ehrenfest's Paradox in the following simplified form. There are two thin rings with radii $R_1$ and $R_2$ (and $R_1 = R_2$). The second ring is speed-up to a linear velocity $V$ by some external device. Let the observers in the laboratory frame measure circumferences of these rings ($L_1$ and $L_2$) in the framework of the relativistic theory methods. There are two hypotheses that present the results of these measurements. Both of them have been considered by many (sometimes famous) authors. For simplicity, we shall refer to two papers, [3] и [4]. According to [4], the circumference of a rotating ring $L^/$ in the rotating frame (rotating with linear velocity $V$ at the radius $R_2$) is $L^/ = \dfrac{1}{\sqrt{1-\dfrac{V^2}{c^2}}} \cdot L_2 = \gamma \cdot L_2$, and $R_1 = R_2$. We shall name it "the condition (1)".

Let's assume that the measured circumference by the observer in the rotating reference frame does not change due to rotation and is equal to the initial length of the non-rotating ring $L_1$. Then the laboratory observers would draw a conclusion, that $L_2 = \gamma^{-1} \cdot L_1$. We shall name it "the hypothesis 1". These are the arguments which were posed by P. Ehrenfest to A. Einstein. It is necessary to emphasize, that the geometry of rotation ring points in the laboratory frame is a Non-Euclidean geometry in the case of the hypothesis 1.

The authors [3], as a result of the analysis of a metric tensor for a rotating frame, have come to a following conclusion. The condition (1) is fulfilled but the rotating observers see the real increase of the circumference in the form of $L^/ = \gamma \cdot L_1$. The laboratory observers would see a relativistic contraction of a rotating ring in the form of $L_2 = \gamma^{-1} \cdot L^/ = \gamma^{-1}\gamma \cdot L_1 = L_1$. We shall name it "the hypothesis 2".

Unfortunately, it is practically impossible to resolve the Ehrenfest's Paradox by observing the real rotating disks or rings. The reason is clear. Centrifugal forces lead to essential deformation of the real rings. Thus, it is impossible to measure very small relativistic effects against the background of that deformation at accessible rotation velocities.

Recently, in [1, 5] the effect of relativistic contraction of an «electron ring» circumference in steady state tokamak plasma rotating in toroidal direction with current velocity $V_e(r)$ has been analyzed. Let $r$ be the radius of a tokamak magnetic surface. The minor tokamak radius $a$ was assumed to be much less than the major radius $R$, where $\frac{a}{R} \ll 1$, and electron toroidal rotation velocity was assumed to be moderate so that it would be possible to exclude centrifugal forces in the momentum balance of a plasma [6]. The toroidal rotation velocity of «the ion ring» $V_i(r)$, as a rule, is much less than the toroidal rotation velocity of the «electron ring» which is known from experiments. It was supposed that at the initial moment (with no current) plasma is created from neutral gas (hydrogen or deuterium), the electron density $n_e^0(r)$ and the ion density $n_i^0(r)$ are equal, and the full number of electrons and ions does not vary during a discharge. Electrons and ions can move and can be redistributed in the minor radius direction of a tokamak plasma after occurrence of the current. The author notes that the maximum of the experimentally measured radial electric field $E_r(r)$ in tokamak corresponds to the occurrence in plasma a small difference between electron density $n_e(r)$ and ion density $n_i(r)$ in the laboratory frame, of the order of $\frac{|n_i(r) - n_e(r)|^{max}}{n_e(r)} \approx (\frac{V_e(r)}{c})^2$; and for the ohmic modes: $\frac{|n_i(r) - n_e(r)|}{n_e(r)} \approx (\frac{V_i(r)}{c})^2$. We shall refer to it as "the condition (2)".

Let us assume, that all electron density and ion density, ion toroidal rotation velocity and electron rotation velocity are constant and do not depend on the tokamak minor radius $r$. In this case an initial electron density (before plasma current) is $n_e^0$ and ion density is $n_i^0$, where $n_e^0 = n_i^0$.

Therefore, one can say that, actually, there are two thin rings (the electron ring and the ion ring), originally having the same circumference $L^{electron} = 2\pi R = L^{ion}$, which are brought to different toroidal rotation velocities, $V_e$ and $V_i$, where $V_i \ll V_e$. The situation is similar to the one considered above in the context of the Ehrenfest's Paradox. In the frame of the hypothesis 1, circumferences of the rotating electron ring $L_{rot}^{electron}$ and the rotating ion ring $L_{rot}^{ion}$ are different in the laboratory frame. In the case of the constant number of electrons and ions, the electron density $n_e$ and the ion density $n_i$ in laboratory frame are changed by relativistic effect. The charge density $\rho = |e|(n_i - n_e) \neq 0$ appears, and the radial electric field $E_r$ is created in a tokamak plasma. Measurement of the part of the electric field which can arise in the frame of the hypothesis 1, is much easier, than the investigation of the deformations of a rotating rigid ring. The reason is as follows: on one hand, the current electron velocity can reach hundreds km/s, on the other hand, possible deformations of the "electron ring» due to centrifugal force lead only to the occurrence of dipole components in electric field associated with minor change of radius of the rotating ring. Relativistic contraction of the ring circumference without change of radius and conservation of full electron number (in the frame of the hypothesis 1) can lead to the occurrence of a monopole component in electric field which is relatively easy to measure, as it will be shown below.

Following [1], it is possible to show, that the rotation may create a density of charges $\rho$ in tokamak plasma. If we ignore higher-order terms in $\frac{V^2}{c^2}$ expansion, we can write:

$$\rho = \frac{|e|n_i'}{\sqrt{1-\frac{V_i^2}{c^2}}} - \frac{|e|n_e'}{\sqrt{1-\frac{V_e^2}{c^2}}} \cong |e|n_i' \cdot (1+\frac{1}{2}\frac{V_i^2}{c^2}) - |e|n_e' \cdot (1+\frac{1}{2}\frac{V_e^2}{c^2}) \cong$$

$$\cong -\frac{(|e| \cdot n_e \cdot (V_i - V_e))^2}{2 \cdot c^2 \cdot |e|n_e} + \frac{(|e| \cdot n_e \cdot (V_i - V_e)) \cdot V_i}{c^2} + |e|(n_i' - n_e') = \quad (3).$$

$$= -\frac{j^2}{2 \cdot c^2 \cdot |e|n_e} + \frac{j \cdot V_i}{c^2} + |e|(n_i' - n_e')$$

Where $n_e'$ and $n_i'$ are: the electron density in the rotating frame with the velocity $V_e$ and the ion density in the rotating frame with the velocity $V_i$. It is the general equation with yet undetermined term $|e|(n_i' - n_e')$.

Let's go back to the Ehrenfest's Paradox. In the case of the hypothesis 2, $L_{rot}^{electron} \equiv L^{electron} \equiv L^{ion} \equiv L_{rot}^{ion}$ and $n_e' = n_e^0 \cdot \sqrt{1-\frac{V_e^2}{c^2}}$, $n_i' = n_i^0 \cdot \sqrt{1-\frac{V_i^2}{c^2}}$. The eq. (3) has the form:

$$\rho = 0 \quad (4).$$

In the case of the hypothesis 1, $L_{rot}^{electron} = L^{electron} \cdot \sqrt{1-\frac{V_e^2}{c^2}} \neq L_{rot}^{ion} = L^{ion} \cdot \sqrt{1-\frac{V_i^2}{c^2}}$ and $n_i' = n_i^0 = n_e^0 = n_e'$. The eq. (3) has the form:

$$\rho \cong -\frac{j^2}{2 \cdot c^2 \cdot |e|n_e} + \frac{j \cdot V_i}{c^2} \quad (5).$$

Change of the charge density in this case is only associated with relativistic change of the denominator in the expression for density. Let us note that $\rho$ depends on parameters measured in the laboratory frame: the current density $j$, the electron density $n_e$ and the ion toroidal rotation velocity $V_i$.

So we have calculated the charge density $\rho$ in the each point inside a tokamak non moving chamber in laboratory frame. One can "forget" about the particular nature of that charge density $\rho$ related to Non-Euclidean geometry of rotation electron (ion) ring points of the tokamak plasma in the laboratory frame and one can use the Poisson equation with $\rho$ taken from eq. (5) to calculate the electrostatic radial electric field in the tokamak plasma. Hence, $E_r(r)$ in tokamak plasma is created by two relativistic terms in the density of charges $\rho$, eq. (5), appeared in the laboratory frame.

In case of a real tokamak, plasma parameters depend on minor radius of magnetic surfaces. Consideration of such dependence for the purpose of calculation of tokamak plasma charge density is shown in [1] in detail for $\frac{a}{R} \ll 1$. The principal point here is the consideration of each nested magnetic surface with plasma just in a thin hollow plasma ring.

Following [1], we can rewrite the eq. (3) so as to take into account the processes of redistribution of electrons and ions on the minor radius by plasma diffusion (convection):

$$\rho(r) \cong -\frac{j^2(r)}{2 \cdot c^2 \cdot |e| n_e(r)} + \frac{j(r) \cdot V_i(r)}{c^2} + |e|(n_i^/(r) - n_e^/(r)) + |e|(n_i^d(r) - n_e^d(r)) \quad (6),$$

where $j = |e| \cdot n_e(r) \cdot (V_i(r) - V_e(r))$. Due to the electron and ion diffusion (change of the numerator in the expression for density) additional volume charge densities in plasma can arise, and it can be expressed by the term $|e|(n_i^d(r) - n_e^d(r))$ in (6).

In the case of the hypothesis 2, we can rewrite the eq. (4) in the following form:

$$\rho(r) \cong |e|(n_i^d(r) - n_e^d(r)) \qquad (7).$$

Hence, only the diffusion (convection) of ions and electrons could be created $E_r(r)$ in tokamak plasma.

In the case of the hypothesis 1, the eq. (5) has the new form:

$$\rho(r) \cong -\frac{j^2(r)}{2 \cdot c^2 \cdot |e| n_e(r)} + \frac{j(r) \cdot V_i(r)}{c^2} + |e|(n_i^d(r) - n_e^d(r)) \qquad (8).$$

The eq. (8) has two relativistic terms, and, by the way, the "condition (2)" is not casual coincidence in this case. Let us emphasize again that $\rho(r)$ depends on the plasma parameters measured in laboratory frame: the current density $j(r)$, the electron density $n_e(r)$, the ion toroidal rotation velocity $V_i(r)$ and the diffusion (convection) term. So we have calculated the charge density $\rho$ in the each point inside tokamak non-moving chamber in the laboratory frame, eq. (8). As emphasized above one can "forget" about the particular nature of that charge density $\rho$ related partially to Non-Euclidean geometry of rotation electron (ion) ring points of the tokamak plasma in the laboratory frame and one can use the Poisson equation with $\rho$ taken from eq. (8) to calculate the electrostatic radial electric field in the tokamak plasma. In our consideration the diffusion (convection) term is not determined. We can mention about one integral property of the diffusion (convention) term, which is a consequence of the physical assumption that the total number of electrons and ions does not vary during a discharge inside the tokamak chamber. It is:

$$\int_{V_{ch}} |e|(n_i^d(r) - n_e^d(r))dV = 0 \qquad (9),$$

where $V_{ch}$ is the volume of the toroidal tokamak chamber. The diffusion (convention) term can be determined in the frame of a different approach, see [1].

The author of the article [1] has compared results based, in fact, on the eq. (8) with the results of real tokamak experiments, and was inclined to believe that the hypothesis 1, most likely, worked.

Let's emphasize that the full number of electrons and ions in the tokamak volume is remained constant during the discharge. In this sense, the given plasma object is closed. But in order for the current in real plasma to occur, the externally initiated an inductive toroidal electric field is necessary. In this general sense, the considered system is not closed.

**The tokamak experiment related to the Ehrenfest's Paradox**

Having accepted the hypothesis 1, we have seen that plasma current creates relativistic volume charge density equal to $-\dfrac{j^2}{2 \cdot c^2 \cdot |e| n_e}$. The second relativistic right-hand term of eq. (8) for plasma usually is more than five times smaller than $-\dfrac{j^2}{2 \cdot c^2 \cdot |e| n_e}$. The third term is the symmetrical redistribution of charges by the diffusion (convection) along the minor radius in a plasma chamber. Thus $-\dfrac{j^2}{2 \cdot c^2 \cdot |e| n_e}$ can be crucial in the creation of $E_r(r)$, especially at the beginning of a discharge, and if plasma has modulated

current. For a tokamak plasma contained in metallic chamber, $E_r(r)$ can modify the chamber electric potential

$$\Delta\varphi \sim \frac{\int_{V_{ch}} \rho(r)dV}{C} \approx \frac{\int_{V_{ch}} (-\frac{j^2(r)}{2\cdot c^2 \cdot |e|n_e(r)} + |e|(n_i^d(r) - n_e^d(r)))dV}{C} = \frac{\int_{V_{ch}} (-\frac{j^2(r)}{2\cdot c^2 \cdot |e|n_e(r)})dV}{C}$$

with respect to the ground; see the eq. (9). Chamber electric potential is proportional in this case to the volume of plasma, the averaged value of $-\frac{j^2}{2\cdot c^2 \cdot |e|n_e}$, the electric capacitance of close metallic tokamak chamber C and relates to chamber RC time.

If one wants to measure the potential of the tokamak chamber $\Delta\varphi(t)$ during the discharge, one can expect two options. In the case of the hypothesis 1 - the potential of the chamber will change proportional to $\frac{\int_{V_{ch}} (-\frac{j^2(r)}{2\cdot c^2 \cdot |e|n_e(r)})dV}{C}$, in the case of the hypothesis 2 - no change of $\Delta\varphi(t)$ due to the plasma current will occur, i.e. $\Delta\varphi(t) = 0$, see the eq. (9).

Thus, measurement of tokamak chamber potential $\Delta\varphi(t)$ during discharges could resolve the Ehrenfest's Paradox in principle.

The first series of special experiments for electric potential measurements in several tokamak chamber points was carried out at T-11M tokamak (main plasma parameters in presented shots were: deuterium plasma, the average steady-state electron density $<n_e> \sim 10^{13}$ cm$^{-3}$, the plasma current Ip ~50 kA, $r$ =20 cm, $R$ =70 cm) with modulated current [7]. The example of the typical measurement is shown on Fig.1. For the purpose of calculation of the theoretical dependence (dashed curve on Fig. 1) we have used: a) experimental data for plasma current and electron density; b) experimental chamber resistivity R = ~4 MOm; c) experimental chamber RC ~ 2.5 ms. Electron density diagnostics did not give us adequate information for few milliseconds in the beginning of discharge. We have extrapolated the electron density growth during the first ~8 ms by a linear function.

One can see satisfactory coincidence of theoretical calculation results based on the hypothesis 1 with the experimental results.

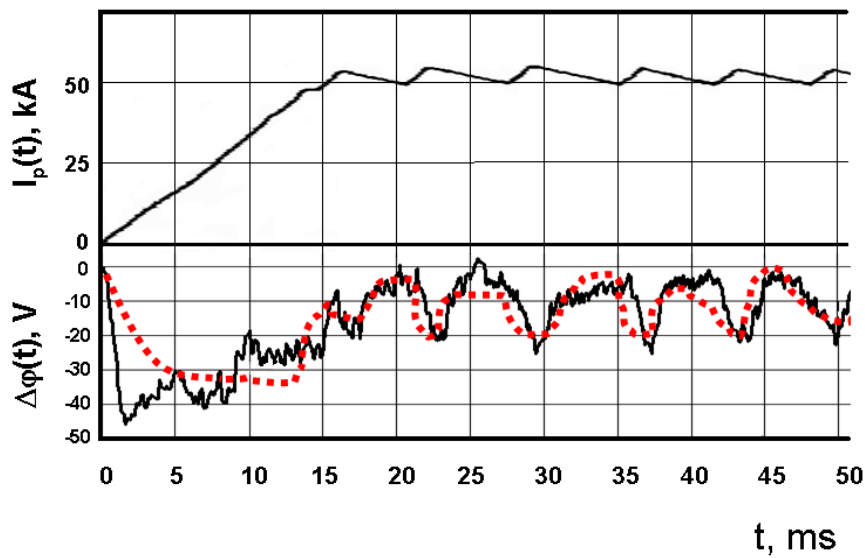

**Fig.1. Time dependence of plasma current $I_p(t)$ and the tokamak chamber electric potential $\Delta\varphi(t)$ during the discharge, T-11M tokamak shot #024825 [7]. Solid curve is the experimental dependence of $\Delta\varphi(t)$; dashed curve is the theoretical dependence.**

**Conclusion**

The main conclusions of the presented investigation are:

1) tokamak plasma can be a tool for the research of possible physical consequences of the Ehrenfest's Paradox. Measurement of tokamak chamber potential $\Delta\varphi(t)$ with respect to the ground during discharges could resolve that Paradox in principle. The plasma created from initially neutral gas inside metallic tokamak chamber can affect to $\Delta\varphi(t)$ by two following ways.

a) The particular effect is $\Delta\varphi(t) \sim \dfrac{\int_{V_{ch}}(-\dfrac{j^2(r)}{2\cdot c^2 \cdot |e| n_e(r)})dV}{C}$. In this case the Ehrenfest's Paradox should be resolved in the frame of the hypothesis 1.

b) The most expected effect is $\Delta\varphi(t)=0$. In this case the Ehrenfest's Paradox should be resolved in the frame of the hypothesis 2.

2) on one hand, good agreement of theoretical results, based on eq. (8), with experimental results, presented in [1], and, on the other hand, available experiments of $\Delta\varphi(t)$ measurements described above, show, that the Ehrenfest's Paradox could be resolved in the frame of the hypothesis 1 much more probably than in the frame of the hypothesis 2.

The second conclusion has the important consequence. The geometry of rotation electron (ion) ring points of the tokamak plasma in the laboratory frame should be a Non-Euclidean geometry.

Though the presented results agree sufficiently well with the hypothesis 1, but there is still one outstanding question. Let's assume, for the sake of simplicity, that $n_e^0 = n_i^0$ in all physical points of the toroidal tokamak chamber at the initial time. Upon the occurrence of the current, the "electron ring" obtains the toroidal rotation velocity $V_e$, and the "ion ring" remains motionless. In the frame of the hypothesis 1, the laboratory observers measure new the electron density in all points of the chamber, namely $n_e^1 = n_e^0 \cdot \frac{1}{\sqrt{1-(\frac{V_e}{c})^2}}$. Sum up $n_e^1 - n_i^0$ on all points inside of the chamber, the laboratory observers, following classical logic, should draw a paradoxical conclusion that the area inside of the chamber appears charged. But the quantity of electrons inside of the chamber has not changed. In the frame of the hypothesis 2, $n_e^1 - n_i^0 \equiv 0$. The hypothesis 2 agrees with the presented experiment rather poorly. The hypothesis 1 agree with the experimental results, but it contradicts to classical representations of electrodynamics [4], namely with the definition of local charge density and with the definition of the full charge in closed volume. In the frame of the hypothesis 1, one has to assume, that the integral $\int e \cdot n_e^1 dV$ is not equal to the total number of electrons $e \cdot N_e$ in the closed volume within the tokamak chamber. Consequently, using the integral form of the Poisson equation demands particular consideration.

So, at least there is one difficult question: "How is the integral form of the Poisson equation necessary to interpret in the frame of the hypothesis 1, when $\int e \cdot n_e^1 dV$ is not equal to full electron number $e \cdot N_e$ in the closed laboratory volume?"

**Acknowledgements**

I am deeply in debt to Dr. Kusnetsov E. who helped in carrying out experiments and measurements at T-11M tokamak.